# Spin-filter tunnel junction with matched Fermi surfaces

T. Harada[1,†], I. Ohkubo[1, a),*], M. Lippmaa[2], Y. Sakurai[1], Y. Matsumoto[3], S. Muto[4], H. Koinuma[5,6], and M. Oshima[1]

[1]*Department of Applied Chemistry, The University of Tokyo, 7-3-1, Hongo, Bunkyo-ku, Tokyo, 113-8656, Japan*
[2]*Institute for Solid State Physics, The University of Tokyo, 5-1-5, Kashiwanoha, Kashiwa, Chiba, 277-8581, Japan*
[3]*Materials and Structures Laboratory, Tokyo Institute of Technology, 4259, Nagatsuta, Yokohama, 226-8503, Japan*
[4]*Department of Materials, Physics, and Energy Engineering, Graduate School of Engineering, Nagoya University, Nagoya 464-8603, Japan*
[5]*Graduate School of Frontier Sciences, The University of Tokyo, 5-1-5, Kashiwanoha, Kashiwa, Chiba 277-8568, Japan*
[6]*Dept. of cogno-Mechatronic Engineering, Pusan National University, Korea*
a) Author to whom correspondence should be adressed.
Electronic mail: **ohkubo@sr.t.u-tokyo.ac.jp**

[†] Now at Institute for Solid State Physics, The University of Tokyo, 5-1-5, Kashiwanoha, Kashiwa, Chiba 277-8581, Japan
[*] Now at National Institute for Materials Science, 1-1 Namiki Tsukuba-shi, Ibaraki 305-0044, Japan
E-mail: OHKUBO.Isao@nims.go.jp

Efficient injection of spin-polarized current into a semiconductor is a basic prerequisite for building semiconductor-based spintronic devices. Here, we use inelastic electron tunneling spectroscopy to show that the efficiency of spin-filter-type spin injectors is limited by spin scattering of the tunneling electrons. By matching the Fermi-surface shapes of the current injection source and target electrode material, spin injection efficiency can be significantly increased in epitaxial ferromagnetic insulator tunnel junctions. Our results demonstrate that not only structural but also Fermi-surface matching is important to suppress scattering processes in spintronic devices. [DOI:     ]

A common component in any spintronic devices is a spin injector that converts conventional electric current into a flow of spins or a spin-polarized current [1-4]. Conventional semiconductor spintronic devices are based on spin injectors that use a ferromagnetic metal (FM) electrode. The electronic structure of the ferromagnetic metal spin injector electrode is inevitably different from that of the semiconductor. The injected spin-polarized electrons therefore need to change their in-plane momentum and possibly the orbital symmetry to enter empty states in the semiconductor. Such adjustment is possible if the electrons crossing the metal-semiconductor interface are involved in scattering by phonons, magnons, crystal defects, etc. Scattering by magnons or defects can be accompanied by an electron spin flip [5], resulting in a drop of spin injection efficiency.

Here, we show that it is possible to achieve high spin injection efficiency in a spin-filter tunnel junction (SFTJ) by matching the Fermi-surface shapes of the spin injection source and target materials. Spin injection through a spin filter tunnel barrier has been studied because it can avoid the impedance mismatch problem that exists between a metallic electrode and a semiconductor [6, 7]. Another advantage, which is the focus of this work, is the availability of a wider choice of current injector electrode materials. In SFTJs, the current injector elec-trode does not need to be spin-polarized or even ferromagnetic, as the spin selectivity of the injected electrons is facilitated by the exchange-split ferromagnetic insulator tunnel barrier. Matching the Fermi surface shapes of the current injector electrode and target materials mitigates the effects of spin scattering due to the availability of a direct tunneling channel.

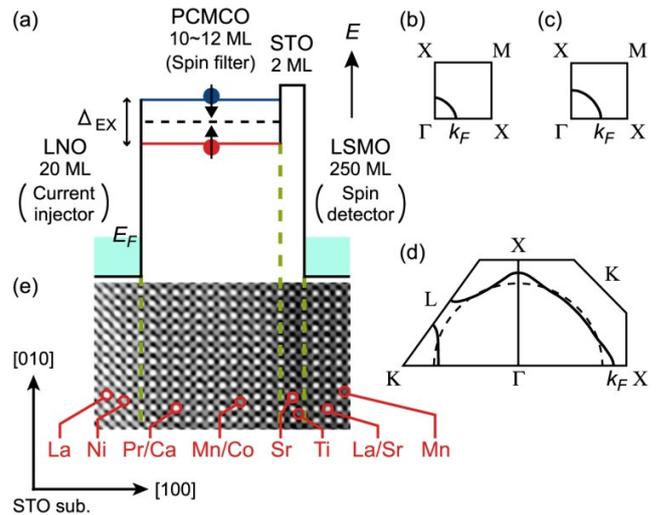

FIG. 1. Schematic illustration of the spin filter tunnel junction operation. (a) Simplified band diagram of a SFTJ device. (b) Cross-sectional Fermi surface plots for LNO [44, 45], (c) La$_{2/3}$Sr$_{1/3}$MnO$_3$ [25], and (d) Au [46]. The sizes of the illustrations are proportional to the size of the first Brillouin zone, allowing comparison of the Fermi momenta k$_F$. The Fermi surface of La$_{2/3}$Sr$_{1/3}$MnO$_3$ depicted in (c) is slightly larger than that of the bottom electrode material La$_{0.6}$Sr$_{0.4}$MnO$_3$. (e) Cross-sectional HAADF-STEM image of a device.

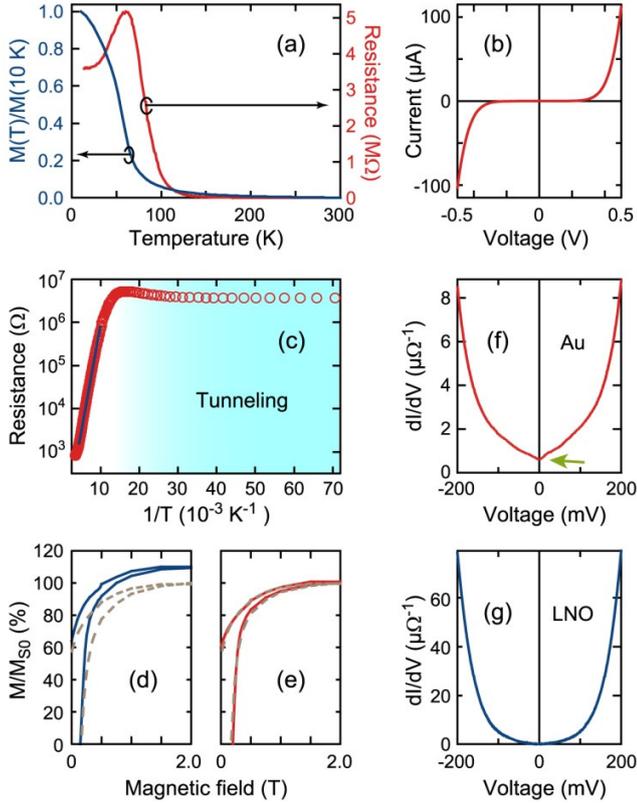

FIG. 2. Transport and magnetic properties of SFTJs. (a) Resistance of a Au/PCMCO (y=0) (12 ML)/LSMO junction (red line) and magnetization of a PCMCO (y=0) film (blue) as a function of temperature. The junction resistance was measured by applying a bias voltage of ±100 mV without an external magnetic field. The magnetization is normalized to the 10 K value. (b) A current-voltage characteristic of a Au/PCMCO (y=0) (12 ML)/LSMO junction at 13 K. (c) Logarithmic plot of the junction resistance as a function of reciprocal temperature. The blue line is a Poole-Frenkel fit for the high temperature region (>100 K). Tunneling conduction is dominant in the highlighted region. (d), (e) Comparison of M-H curves of PCMCO (15 ML) films grown on STO substrates before (gray dashed line) and after depositing either epi-LNO ((d), blue line) or poly-Au ((e), red line) electrodes, measured at 10 K [47]. The vertical axes are normalized to the saturation magnetization of the PCMCO film ($M_{S0}$). The dI/dV spectra of (f) poly-Au/PCMCO (y=0.2) (10 ML)/STO (2 ML)/LSMO junction and (g) LNO/PCMCO (y=0.2) (10 ML)/STO (2 ML)/LSMO junction measured at 4 K and 0 T. A zero-bias anomaly is indicated by the green arrow. The dI/dV spectra were measured at zero field after cooling in a magnetic field of 0.8 T.

The operation of a SFTJ is based on spin-dependent tunneling in a ferromagnetic insulator [8-19]. As shown in a simplified band diagram in Fig. 1(a), the energy bands in a ferromagnetic insulator are spin-split due to the presence of an exchange splitting ($\Delta_{EX}$). This results in two spin-dependent tunnel barrier heights, with the up-spin electrons seeing a lower effective barrier than the down-spin electrons and thus preferentially tunneling through the junction, generating a spin-polarized current. In a practical spintronic device, the spins would be injected into a semiconducting material, as in spin transistors or spin light-emitting diodes [20-22]. Scattering processes in a tunnel junction can be investigated using inelastic electron tunneling (IET) spectroscopy.

In this work, our purpose is to develop a high-performance SFTJ by suppressing the effect of scattering. To realize spin detection and Fermi-surface matching at the same time, a suitable set of normal and ferromagnetic metals is employed as electrodes. Spin-polarized current from the spin filter tunnel barrier is detected by ferromagnetic metals as tunnel magnetoresistance (TMR) signals, i.e. TMR = $(R_{AP}-R_P)/R_P$, where $R_P$ and $R_{AP}$ denote the junction resistances in parallel and antiparallel magnetization configurations of the spin filter and spin detector layers.

In the devices studied here (Fig. 1(a)), electrons emitted from a metallic electrode are filtered by a ferromagnetic insulator, $Pr_{0.8}Ca_{0.2}Mn_{1-y}Co_yO_3$ (PCMCO) [23]. The resulting spin-polarized current is injected into a ferromagnetic $La_{0.6}Sr_{0.4}MnO_3$ (LSMO) spin detector layer. To establish structural and Fermi-surface matching between the LSMO spin detector and the injector electrode, a paramagnetic metal, $LaNiO_3$ (LNO), was used. LNO and LSMO have similar Fermi surface shapes and Fermi momenta ($k_F$), as shown in Figs. 1(b) and (c) [24, 25]. The conduction electrons in LNO ($Ni^{3+}$ $3d\ e_g$) and LSMO ($Mn^{3.6+}$ $3d\ e_g$) have the same orbital symmetry, which opens a direct tunneling channel that conserves the in-plane momentum and orbital symmetry. The influence of scattering by phonons, magnons or defects on the total injection current is thus greatly reduced. Moreover, LNO, PCMCO and LSMO are all perovskite-type oxides, and can thus be combined in a lattice-matched epitaxial heterostructure with a very low density of interfacial structural defects that can also act as scattering centers. Reference structures with Au current injector electrodes were also fabricated for comparison to show that a large mismatch in the Fermi-surface sizes of SFTJ electrodes, as illustrated in Fig. 1(d), leads to increased contributions of scattering. A thin nonmagnetic insulator (NI) $SrTiO_3$ (STO) spacer layer was inserted between the PCMCO and LSMO layers to magnetically decouple the two ferromagnetic materials.

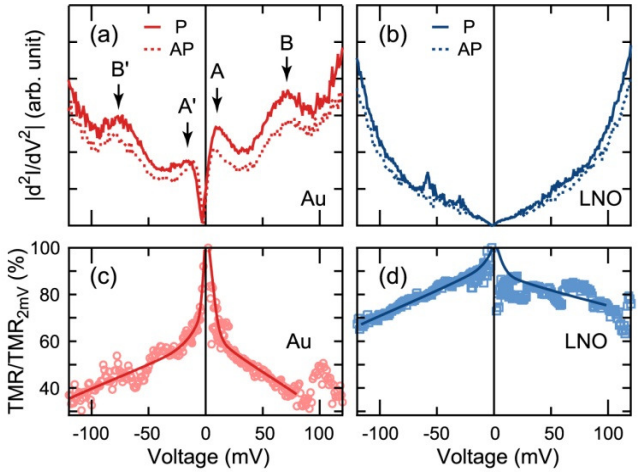

FIG. 3. Suppression of electron scattering using epitaxial current injectors. IET spectra of (a) poly-Au/PCMCO (y=0.2) (10 ML)/STO (2 ML)/LSMO junction and (b) LNO/PCMCO (y=0.2) (10 ML)/STO (2 ML)/LSMO junction at 4 K. The solid (dotted) lines are measured in P (AP) state under 0.8 T (0.05 T). Peak positions are indicated by black arrows. Bias voltage dependence of TMR at 4 K in (c) poly-Au/PCMCO (y=0.2) (10 ML)/STO (2 ML)/LSMO and (d) LNO/PCMCO (y=0.2) (10 ML)/STO (2 ML)/LSMO junctions. Lines are visual guides.

Epitaxial LNO/PCMCO/STO/LSMO layered structures were grown on atomically flat STO (001) substrates by pulsed laser deposition. The optimal growth temperature and oxygen pressure were 900 °C and 0.08 Pa for the LSMO bottom electrodes, and 700 °C, 4 Pa for the STO, PCMCO, and LNO layers. The thickness of the PCMCO/STO tunnel barriers was controlled by counting the reflection high-energy electron diffraction specular intensity oscillations. The Au top electrode was deposited by standard vacuum evaporation. The multilayers were patterned into 8×32 μm$^2$ junctions using a conventional photolithography process and post-annealed in air for 24 h to reduce the number of oxygen vacancies. The device bias is defined positive when a positive bias is applied to the LNO or Au top electrode of the tunnel junctions.

A cross-sectional high-angle annular dark-field scanning transmission electron microscopy (HAADF-STEM) image in Fig. 1(e) shows that all interfaces are epitaxial and atomically abrupt. The thickness of each layer is noted in terms of the perovskite unit cell height (1 ML ~0.4 nm).

As shown by the temperature dependence of a Au/PCMCO (12 ML)/LSMO junction resistance (red line) in Fig. 2(a), even very thin (12 ML) PCMCO tunnel barrier films were excellent insulators. The nonlinear current-voltage characteristic in Fig. 2(b) shows that the dominating mode of charge transport through the junction is tunneling at low temperature. This can also be confirmed from the logarithmic plot of the junction resistance shown in Fig. 2(c). At high temperature (>100 K), the temperature dependence of the junction resistance follows the Poole-Frenkel model. At low temperatures, as highlighted in Fig. 2(c), the logarithmic plot of the junction resistance becomes almost flat, supporting the occurrence of tunneling conduction. Below the ferromagnetic transition temperature of PCMCO ($T_C \sim 75$ K), determined from the temperature dependence of magnetization (blue line) in Fig. 2(a), the opening of the exchange splitting gap decreases the tunnel barrier height for the up-spin electrons. This effect is evidenced by a sharp drop in the junction resistance at around the $T_C$ of PCMCO, as seen in a linear plot of the junction resistance in Fig. 2(a) (red line), giving a clear signature of the expected spin filtering effect in the PCMCO layer [26].

In most SFTJs studied so far [12, 13, 16, 19], polycrystalline Au (poly-Au) electrodes have been used as the injector electrode. The role of the current injector electrode material in determining the spin injection efficiency in SFTJs has not been considered. However, the tunneling regime in such junctions is not determined purely by the tunnel barrier characteristics, but also by the electronic structures of the electrodes, as has been demonstrated in single-crystalline Fe/MgO/Fe TMR devices [27].

The structural advantage of using an epitaxial oxide electrode can be seen by the increase of the saturation magnetization in the topmost 1~2 ML of PCMCO after the deposition of an epitaxial LNO (epi-LNO) layer on a PCMCO surface, as shown in Fig. 2(d). It is known that in perovskite manganites, ferromagnetism can be degraded close to interfaces [28]. By fabricating a fully epitaxial layered structure, ferromagnetism of the topmost PCMCO layer can be recovered. However, no recovery of ferromagnetism was observed when a poly-Au electrode was deposited. This can be seen in the identical saturation magnetizations of a PCMCO layer before and after the deposition of a poly-Au electrode in Fig. 2(e), indicating that there are fewer scattering sources, such as interfacial defects and misaligned spins [29, 30], in the epi-LNO junction than in the poly-Au junction.

Electron tunneling spectroscopy was used to determine the role of scattering in the tunneling conduction of the SFTJs. In this technique, the differential conductance (dI/dV) of a junction is plotted against the junction bias, which can be obtained by measuring the AC component of the tunneling current. When a poly-Au electrode was used as a current injector, a distinct dip structure appeared at close to zero bias (Fig. 2(f), green arrow).

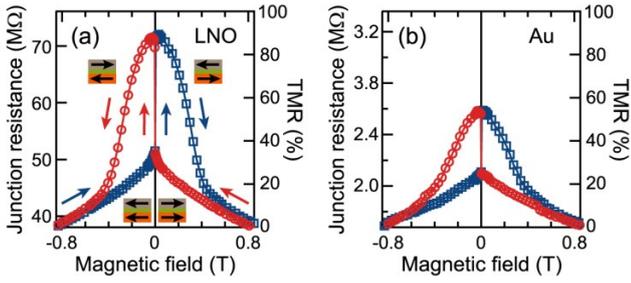

FIG. 4. Comparison of TMR curves of (a) LNO/PCMCO (y=0.2) (10 ML)/STO (2 ML)/LSMO and (b) poly-Au/PCMCO (y=0.2) (10 ML)/STO (2 ML)/LSMO junctions measured at 4 K under a bias voltage of -5 mV. The scan direction is indicated by red and blue arrows in (a).

This zero-bias anomaly is known to be caused by inelastic scattering of tunneling electrons [31]. The advantage of using an epitaxial LNO injector electrode becomes clear when the Au junction data is compared with the dI/dV characteristic obtained from an epi-LNO device, shown in Fig. 2(g). There is no zero-bias anomaly or any other dip structures that could be assigned to inelastic scattering processes, indicating that electrons can pass through the spin-filter barrier by direct tunneling.

Further investigation of the scattering process is possible using IET spectra, i.e., $d^2I/dV^2$ plots obtained from the second-order harmonics of the AC tunneling current, as shown in Figs. 3(a) and 3(b). Broad peaks were observed in the poly-Au junction spectrum (Fig. 3(a)) at around ±70 mV (B, B'), which are composed of two sets of peaks at ±59 mV and ±78 mV. These peaks coincide with phonon excitations (±59 mV, ±76 mV) that have been observed by Raman spectroscopy in an epitaxial $Pr_{0.4}Ca_{0.6}MnO_3$ thin film [32, 33]. This shows that the tunneling electrons were inelastically scattered by phonons inside the PCMCO layer. These phonon peaks were suppressed in the epi-LNO junction, as shown in Fig. 3(b), indicating that electrons were tunneling between the epi-LNO and LSMO electrodes without being scattered by phonons in the tunnel barrier. The suppressed contribution of the electron-phonon scattering is caused by the similarity of the Fermi-surface shapes of LNO and LSMO that determine the initial and final positions of the tunneling electrons in $k$-space (Fig. 1(b) and (c)). If the Fermi-surface shapes of the electrodes are quite different, as in the case of the poly-Au junction, tunneling electrons are likely to be involved in inelastic scattering processes in order to adjust their positions in $k$-space during tunneling by emitting or absorbing phonons (Fig. 1(c) and (d)) [34, 35]. The similarity of the Fermi-surface shapes of LNO and LSMO would allow electrons in LNO to directly tunnel through the PCMCO barrier into the empty states in LSMO without being scattered by phonons [24, 25].

The peaks at approximately ±10 mV (A, A') in the poly-Au device IET spectrum are related to the zero-bias anomaly observed in the dI/dV spectrum. Peaks in this region are often attributed to magnons and defect scattering [36, 37], which can cause electron spin flips in a ferromagnetic junction [5]. The (A, A') peaks can thus be assigned to the appearance of magnetic scattering of tunneling electrons in the poly-Au junction. This interpretation was verified by comparing the bias voltage dependence of TMR of the poly-Au and LNO junctions, as discussed later. As shown in Fig. 3(b), the magnetic scattering peaks are also suppressed by using an epi-LNO electrode, due to the availability of the direct tunneling channel. The intensity of the magnetic scattering peaks did not show a strong dependence on the relative magnetic orientation, either parallel (P) or antiparallel (AP), of the PCMCO and LSMO layers. This behavior is quite different from the magnon scattering that has been observed in FM/NI/FM devices [36, 37] and could imply that the effects, origins, and locations of magnetic scattering in SFTJs might be different from those of conventional TMR junctions.

The effect of electrode materials on the spin injection efficiency can be demonstrated by measuring the magnitude of the spin-polarized current with an epitaxial LSMO spin detector. The magnetic field dependences of junction resistances (TMR curves) for representative junctions are shown in Fig. 4. As expected, a systematic increase of the TMR ratio was observed when an epi-LNO top electrode was used instead of gold, as shown by a comparison of plots in Figs. 4(a) and 4(b). This behavior is consistent with the recovery of interface ferromagnetism in PCMCO (Figs. 2(d), (e)) and the suppression of electron scattering peaks (Figs. 3(a), (b)). The difference in the resistances of the Au and LNO junctions may also be affected by structural disorder at the Au/PCMCO interface. The effect of the magnetic scattering can be clearly seen in the bias voltage dependence of the TMR ratio in Figs. 3(c) and 3(d). In a poly-Au junction, the TMR ratio suddenly dropped at approximately ±10 mV, as shown in Fig. 3(c). The voltage region of this sudden TMR drop corresponds to the magnetic scattering peaks (A, A') in Fig. 3(a). By using an epi-LNO current injector, the sharp drop of the TMR ratio in the low bias region was effectively eliminated, as shown in Fig. 3(d). Dominance of the direct tunneling channel and the lack of ferromagnetic deterioration at the interface in the epi-LNO junction can therefore be concluded to be responsible for the dramatic increase of spin injection efficiency. The noise observed in the bias voltage dependences of TMR in Figs. 3 (c) and (d) may be explained by local magnetic fluctuations in the magnetic layers [10]. Further research, such as PCMCO thickness dependence [27, 38] or noise spectroscopy, [39, 40] would be valuable for investigating the coherence of the tunneling electrons.

In summary, we have demonstrated the operation of a high-performance SFTJ based on a PCMCO ferromagnetic insulator tunnel barrier. In order to realize efficient spin injection, it is important to use an epitaxial current injector that has similar crystal and electronic structure to the injection target material. The easiest way to obtain a Fermi-surface-matched electrode is to simply use a same (or similar) material as the injection target material. Recent progress in crystal growth techniques has made it possible to fabricate single crystalline spin filter materials on practically interesting semiconductors, such as Si, GaN, GaAs etc [41-43]. Applying the results of this work to such systems may lead to high-efficiency next-generation spintronic devices based on oxides and traditional semiconductors.

The authors thank Prof. T. Hasegawa for the MPMS measurements. This work was supported by Grants-in-Aid for Scientific Research (Nos. 22015006 and 20047003). T.H. also acknowledges financial support from JSPS. H.K. thanks National Research Foundation of Korea for the support by the World Class University program (grant no. R31-20004).

## References

[1] I. Žutić, J. Fabian, and S. Das Sarma, Rev. Mod. Phys. **76**, 323 (2004).
[2] Y. Ohno, D. K. Young, B. Beschoten, F. Matsukura, H. Ohno, and D. D. Awschalom, Nature **402**, 790 (1999).
[3] P. R. Hammar, B. R. Bennett, M. J. Yang, and M. Johnson, Phys. Rev. Lett. **83**, 203 (1999).
[4] Y. Kajiwara *et al.*, Nature **464**, 262 (2010).
[5] F. Guinea, Phys. Rev. B **58**, 9212 (1998).
[6] G. Schmidt, D. Ferrand, L. W. Molenkamp, A. T. Filip, and B. J. vanWees, Phys. Rev. B **62**, R4790 (2000).
[7] E. I. Rashba, Phys. Rev. B **62**, R16267 (2000).
[8] J. S. Moodera, R. Meservey, and X. Hao, Phys. Rev. Lett. **70**, 853 (1993).
[9] J. S. Moodera, X. Hao, G. A. Gibson, and R. Meservey, Phys. Rev. Lett. **61**, 637 (1988).
[10] P. LeClair, J. K. Ha, H. J. M. Swagten, J. T. Kohlhepp, C. H. van de Vin, and W. J. M. de Jonge, Appl. Phys. Lett. **80**, 625 (2002).
[11] A. Filip, Appl. Phys. Lett. **81**, 1815 (2002).
[12] M. G. Chapline, and S. X. Wang, Phys. Rev. B **74**, 014418 (2006).
[13] U. Lüders *et al.*, Adv. Mater. **18**, 1733 (2006).
[14] B. B. Nelson-Cheeseman, R. V. Chopdekar, L. M. B. Alldredge, J. S. Bettinger, E. Arenholz, and Y. Suzuki, Phys. Rev. B **76**, 220410 (2007).
[15] A. V. Ramos, M.-J. Guittet, J.-B. Moussy, R. Mattana, C. Deranlot, F. Petroff, and C. Gatel, Appl. Phys. Lett. **91**, 122107 (2007).
[16] M. Gajek, M. Bibes, S. Fusil, K. Bouzehouane, J. Fontcuberta, A. Barthélémy, and A. Fert, Nat. Mater. **6**, 296 (2007).
[17] G.-X. Miao, M. Müller, and J. S. Moodera, Phys. Rev. Lett. **102**, 076601 (2009).
[18] Y. K. Takahashi, S. Kasai, T. Furubayashi, S. Mitani, K. Inomata, and K. Hono, Appl. Phys. Lett. **96**, 072512 (2010).
[19] E. Wada, K. Watanabe, Y. Shirahata, M. Itoh, M. Yamaguchi, and T. Taniyama, Appl. Phys. Lett. **96**, 102510 (2010).
[20] S. Datta, Appl. Phys. Lett. **56**, 665 (1990).
[21] J. Nitta, T. Akazaki, H. Takayanagi, and T. Enoki, Phys. Rev. Lett. **78**, 1335 (1997).
[22] R. Fiederling, M. Keim, G. Reuscher, W. Ossau, G. Schmidt, A. Waag, and L. W. Molenkamp, Nature **402**, 787 (1999).
[23] T. Harada, I. Ohkubo, M. Lippmaa, Y. Matsumoto, M. Sumiya, H. Koinuma, and M. Oshima, Phys. Status Solidi RRL **5**, 34 (2011).
[24] N. Hamada, J. Phys. Chem. Solids **54**, 1157 (1993).
[25] J. Krempaský *et al.*, Phys. Rev. B **77**, 165120 (2008).
[26] L. Esaki, P. J. Stiles, and S. von Molnar, Phys. Rev. Lett. **19**, 852 (1967).
[27] S. Yuasa, T. Nagahama, A. Fukushima, Y. Suzuki, and K. Ando, Nat. Mater. **3**, 868 (2004).
[28] J. Choi, J. Zhang, S. H. Liou, P. A. Dowben, and E. W. Plummer, Phys. Rev. B **59**, 13453 (1999).
[29] X. G. Zhang, Y. Wang, and X. F. Han, Phys. Rev. B **77**, 144431 (2008).
[30] Y. Ke, K. Xia, and H. Guo, Phys. Rev. Lett. **100**, 166805 (2008).
[31] E. L. Wolf, *Principles of electron tunneling spectroscopy* (Oxford Univ. Press, New York, 1985).
[32] A. Antonakos, D. Palles, E. Liarokapis, M. Filippi, and W. Prellier, J. Appl. Phys. **104**, 063508 (2008).
[33] See Supplemental Material at [URL] for the peak assignment of the IETS spectrum.
[34] E. O. Kane, J. Appl. Phys. **32**, 83 (1961).
[35] E. Y. Tsymbal, O. N. Mryasov, and P. R. LeClair, J. Phys.: Condens. Matter. **15**, R109 (2003).
[36] Y. Ando, J. Murai, H. Kubota, and T. Miyazaki, J. Appl. Phys. **87**, 5209 (2000).
[37] R. Matsumoto, A. Fukushima, K. Yakushiji, S. Nishioka, T. Nagahama, T. Katayama, Y. Suzuki, K. Ando, and S. Yuasa, Phys. Rev. B **79**, 174436 (2009).
[38] C. W. Miller, J. Magn. Magn. Mater. **321**, 2563 (2009).
[39] R. Guerrero, F. G. Aliev, Y. Tserkovnyak, T. S. Santos, and J. S. Moodera, Phys. Rev. Lett. **97**, 266602 (2006).
[40] Y. M. Blanter, and M. Büttiker, Phys. Rep. **336**, 1 (2000).
[41] S. I. Khartsev, J. H. Kim, and A. M. Grishin, J. Cryst. Growth **284**, 1 (2005).
[42] A. Schmehl *et al.*, Nat. Mater. **6**, 882 (2007).
[43] M. Muller, R. Schreiber, and C. M. Schneider, J. Appl. Phys. **109**, 07C710 (2011).
[44] K. R. Nikolaev *et al.*, Phys. Rev. Lett. **85**, 3728 (2000).
[45] R. Eguchi, A. Chainani, M. Taguchi, M. Matsunami, Y. Ishida, K. Horiba, Y. Senba, H. Ohashi, and S. Shin, Phys. Rev. B **79**, 115122 (2009).
[46] D. J. Roaf, Philos. Trans. R. Soc. London, Ser. A **255**, 135 (1962).
[47] See Supplemental Material at [URL] for the full hysteresis curves.